\documentclass[conference]{IEEEtran}

\usepackage{cite,graphicx,amssymb,amsmath,color,textcomp}
\usepackage[square, comma, sort&compress, numbers]{natbib}

\usepackage{algorithm}
\usepackage{algorithmic}
\usepackage{graphicx}
\usepackage{subfigure}
\usepackage{epsfig}
\usepackage{epstopdf}
\usepackage{mathrsfs}
\usepackage{color}
\usepackage{bm}
\usepackage{epstopdf}
\usepackage{amsfonts,amssymb}
\usepackage{amsthm}
\usepackage{colortbl}

\title{Handover Analysis on High Speed Train with Doppler Frequency Spread}

\author{ Jingxian Liu$^\dag$ and Xiyu Wang$^\dag$\\
$^\dag$ School of Computer and Information Technology, Beijing Jiaotong University, Beijing, China.\\
Corresponding email: 15112093@bjtu.edu.cn.\\
}

\begin{document}

\maketitle                   
\begin{abstract}
This paper investigates the handover performance on high speed train (HST) via different speed under three scenarios, which are viaduct environment, cutting environment and urban area. To provide stable wireless service, we adopt Long Term Evolution (LTE) for high speed railway (HST) system. Moreover, we consider the effect of speed on handover procedure. Interchannel interference (ICI) caused by the Doppler frequency spread is related to the speed of HST, which is considered into the proposed system to evaluate the system handover performance including handover starting time and handover delay. Numerical results show the parameters in the system including SNR, throughput, handover starting time and handover delay. The proposed system provides some insights for practical railway system.\\

\emph{Index Termes}$-$handover performance, high speed train, Doppler frequency spread

\end{abstract}

\section{INTRODUCTION}
In recent years, high speed trains (HST) have becoming more and more popular, and are treated as a convenient and safe public transportation system in the world. During the time period on the HST, the passengers need many wireless mobile services, such as social application, online video, online game and so on \cite{Y. Lu, Y. Lu1, O. B. Karimi}. To provide the stable services for passengers and reduce the economical cost, the Long Term Evolution (LTE) of the Third Generation Partnership Project (3GPP) should be adopted \cite{L. Dai, H. Ghazzai}, which could support at least 200 users in a cell with very high information rates. 

In LTE system, orthogonal frequency-division multiplexing (OFDM) is used for downlink \cite{E}, which have many narrow subchannels divided by the transmission bandwidth. Hence, in this system, the greater Doppler frequency spread caused by the high mobility environment will cause more interchannel interference (ICI), which could degrade the performance of the systems. Some existing works aim to reduce or eliminate the influence of ICI by different channel estimation techniques in special cases, see e.g., \cite{X. Li, X. Ren, J. Lu}. However, for the proposed HSR system, without using new channel estimation techniques, the effect of ICI should be analysed during the handover procedure because it is related to the speed of HST.

In another aspect, as the time cost for passenger is increasing, more higher speed for the HST is necessary. However, it results in much greater Doppler frequency spread and more frequent handover \cite{Y. Yang}. The hard handover in LTE system causes that there exists an interruption for the users, which would happen more frequent because of very high speed. The handover delay exists in the LTE system, and when to start handover is related to whether handover is successful or not. Therefore, high mobility affects not only ICI but also users experience during their communication time on the HST. 

The deployment of high speed railway (HSR) systems is various in different environments \cite{C. Zhang, C. Zhang1}, such as viaduct environment, cutting environment and urban environment. For different environments, channel fading shows very different, which affects the transmitting process. Therefore, taking different environments into consideration for handover procedure is meaningful. In this paper, we consider viaduct environment, cutting environment and urban area for handover analysis. We adopt remote radio head (RRH) with multiple antenna deployed on both sides of the railway and three RRH are assumed to connect to one baseband unit (BBU) with fiber \cite{3GPP}. One kind of environment is assumed between two RRHs. 

The contributions of our work are summarized as follows. \emph{Firstly}, we separately take three scenarios with different channel states into consideration along the railway in the HSR system, which are viaduct environment, cutting environment and urban area. \emph{Secondly}, based on the data which just considers path loss and small-scale channel fading, we add shadow fading and ICI in the system for practical consideration. And we verify that these processed data conforms the original results , so the process is reasonable. \emph{Thirdly}, we analyse the effect of high mobility on the handover scheme. For example, higher speed postpones the handover starting time.  

The rest of this paper is organized as follows. The system model is described in section II. Section III presents the handover procedure and analysis, and section IV provides some numerical results and discussion. Finally, section V concludes the paper.

\begin{figure}[!t]
\centering
\includegraphics[width=0.48\textwidth]{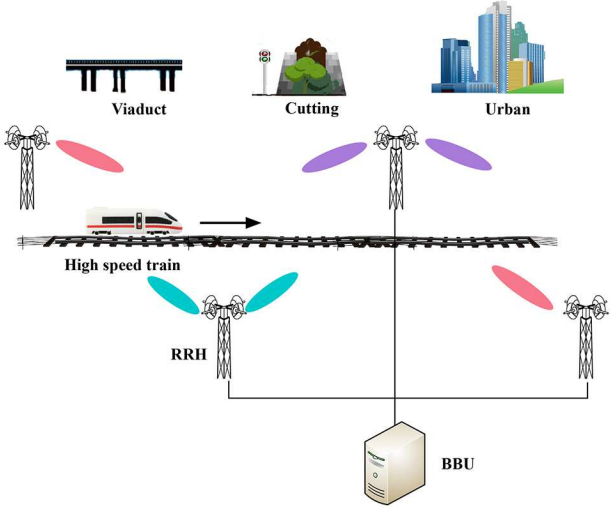}
\caption{Illustration of the proposed railway wireless system.}
\label{fig1} 
\vspace{-3mm}
\end{figure}

\section{SYSTEM MODEL}
As illustrated in Fig. \ref{fig1}, we consider three scenarios in the railway wireless communication system, which are viaduct environment, cutting environment and urban area, and every RRH has two bidirectional beams, which are parallel to the track. In order to analyse the handover performance in different environment, assume that there is one kind of environment between two RRHs. And we assume that one BBU connects to three neighboring RRHs.

\subsection{Channel Description}
In the proposed system, we consider path loss, which relies on the distance between the transmitter and the receiver, and small-scale channel fading. In different environments, their channels are very different from each other in practice. 
\begin{itemize}
\item For urban area, there exists abundant scatterers in the area, which affect the small-scale fading of the channels. 
\item For cutting environment, sometimes there is line-of-sight (LOS) path but sometimes there is non-line-of-sight (NLOS) path. So the channel modeling is not the same all the time.
\item For viaduct environment, there always exists LOS path, which is different from the cutting environment.
\end{itemize}

\subsection{Received Signal Strength (RSS) with Doppler Frequency Spread}
The RSS describes the expectation of the received power, i.e., 
\begin{equation*}
{P_r} = \mathbb{E}\{ PL(d) \zeta {\left| \bm{\mathrm{h}} \right|^2}{P_t}\},
\end{equation*}
where $PL(d)$ denotes the path loss related to the distance $d$ between the transmitter and the receiver, $\zeta \sim \mathcal{CN} (0, \sigma)$ describes the shadow fading, $\bm{\mathrm{h}}$ is the small-scale fading and $P_t$ is the transmit power at RRH. 

High mobility results in the greater Doppler frequency spread, which will cause the ICI in OFDM system. And the Doppler frequency spread has effect on the RSS in dBm units \cite{Z. Liu}, i.e.,   
\begin{equation}\label{RSS}
RSS = 10{\log _{10}}\frac{{{P_r}}}{{{P_r}{P_{ICI}} + 1}},
\end{equation}
where $P_{ICI}$ is the ICI power and the noise power is normalized. 
From (\ref{RSS}), we know that the $P_r$ may decrease with the effect of the $P_{ICI}$. Moreover, the bounds of the $P_{ICI}$ is analyzed in \cite{Ye Li},
shown as follows, 
\begin{equation}
 \begin{cases}
{P_{ICI}} \ge \frac{{{\alpha _1}}}{{12}}{(2\pi {f_d}{T_s})^2} - \frac{{{\alpha _2}}}{{360}}{(2\pi {f_d}{T_s})^4},\\
{P_{ICI}} \le \frac{{{\alpha _1}}}{{12}}{(2\pi {f_d}{T_s})^2},
\end{cases}
\label{PICI}
\end{equation}
where $\alpha _1$ and $\alpha _2$ are in classic model, $f_d$ is the Doppler frequency spread, and $T_s$ is the symbol duration. 

\textcolor[rgb]{0,0,0}{In our work, we just analyse the upper bound of $P_{ICI}$ because we need to find its greatest effect on the $P_r$, which has effect on the handover performance. }

\begin{figure}[!t]
\centering
\includegraphics[width=0.48\textwidth]{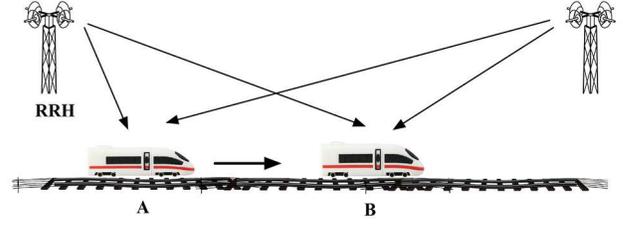}
\caption{High mobility in handover procedure.}
\label{fig2} 
\vspace{-3mm}
\end{figure}

\section{HANDOVER PROCEDURE AND ANALYSIS}
\subsection{Handover Procedure}
We employ the handover scheme dominated by LTE system. In order to analyse the delay and the handover performance, we describe the procedure in detail.
\subsubsection{Filtering} The obtained RSS in handover procedure need to be filtered firstly. And the filtering operations happen in both Layer 1 and Layer 3. For Layer 1, the filtering is shown as follows,
\begin{itemize}
\item Take every sample over 40ms, 
\item Apply a sliding window of 200ms,
\item Transfer the RSS into dB domain,
\item Add lognormal noise with a cutoff,
\item Get the Layer 1 measurements.
\end{itemize}
For Layer 3, the filtering is shown as follows,
\begin{itemize}
\item Run a recursive filter over the Layer 1 measurements, 
\item Get the Layer 3 measurements.
\end{itemize}

\subsubsection{Event A3 Measurement} This event will trigger the handover, which refers to the RSS from target cell (i.e., $RSS_{tar}$) is larger than the RSS from serving cell (i.e., $RSS_{ser}$) by a hysteresis margin $H_0$ over a time period $TTT$, which can be given by 
\begin{equation}\label{RSS}
RS{S_{tar}} - RS{S_{ser}} \ge H_0,
\end{equation}
where $H_0$ includes several parameters (i.e., offset, ofn, ofs and CIO).

\subsubsection{Handover Report} The user equipment (UE) would send a measurement report and check the uplink signal noise ratio (SNR) in the serving cell whether the measurement report goes through. If SNR is too small (e.g., -10dB), there is no need for further action.   

\subsubsection{Handover Preparation} There needs a delay to negotiate between serving cell and target cell to ensure that the target cell gets the UE context. And then it would send handover request ACK to the serving cell.

\subsubsection{Handover Command} In this phase, the serving cell sends the handover command. And check the downlink SNR in the serving cell whether the measurement report goes through. If SNR is too small (e.g., -10dB), there is also no need for further action. 

\subsubsection{Random Access Channel (RACH)}
\begin{itemize}
\item Introduce a delay for system information block (SIB) reading and RACH procedure, 
\item The UE send RACH, and check the uplink SNR and down SNR whether the RACH procedure can be successful. The handover would be failed If the SNR is too small, and then re-establishment procedure would start. If not, the handover would be successful and the UE is now connected to the target cell.
\end{itemize}


After the above phases, if the handover is failed, there will be Radio Link Failure (RLF) part and re-establishment part, which are not described in detail.

\subsection{Analysis}
After the event A3 measurement, the UEs would send a measurement report to the serving cell, and then they move with the HST. So when the serving cell is ready to execute handover, its handover position is not the position where sent the measurement report. As illustrated in Fig. \ref{fig2}, the UEs send the measurement report at position A, but until the HST moves at position B, the UEs receive the handover command.        
We assume that the difference between the RSS from the serving cell (i.e., $RSS_{tar}^{(A)}$) and the RSS from the target cell (i.e., $RSS_{ser}^{(A)})$ at position A is $H_a$, which can be written as
\begin{equation}\label{A}
 RSS_{tar}^{(A)} - RSS_{ser}^{(A)} = {H_a}.
\end{equation}
Similarly, we assume that the difference between the RSS from the serving cell (i.e., $RSS_{tar}^{(B)}$) and the RSS from the target cell (i.e., $RSS_{ser}^{(B)})$ at position B is $H_b$, which can be written as
\begin{equation}\label{B}
 RSS_{tar}^{(B)} - RSS_{ser}^{(B)} = {H_b}.
\end{equation}
The running direction of the HST is from position A to position B, so it must satisfy that $H_a < H_b$.

Based on the above analysis, the relationship among $H_0$ in (\ref{RSS}), $H_a$ in (\ref{A}) and $H_b$ in (\ref{B}) are divided into three categories.
\begin{itemize}
\item If they satisfy
\begin{equation}\label{h1}
{H_a} < {H_b} < {H_0},
\end{equation}
then the handover would not happen because the event A3 is not triggered.
\item If they satisfy
\begin{equation}\label{h2}
{H_0} < {H_a} < {H_b},
\end{equation}
then the handover would happen at position B because the event A3 is already triggered.
\item If they satisfy
\begin{equation}\label{h3}
{H_a} < {H_0} < {H_b},
\end{equation}
then the handover would not happen at position B because the event A3 is not triggered.
\end{itemize}

In fact, (\ref{h1}) and (\ref{h2}) mean that the starting time of handover execution is not affected by mobility. However, from (\ref{h3}), we know that the handover execution would postpone to happen.

\begin{figure}[!t]
\centering
\includegraphics[width=3.5cm, height=2.8cm]{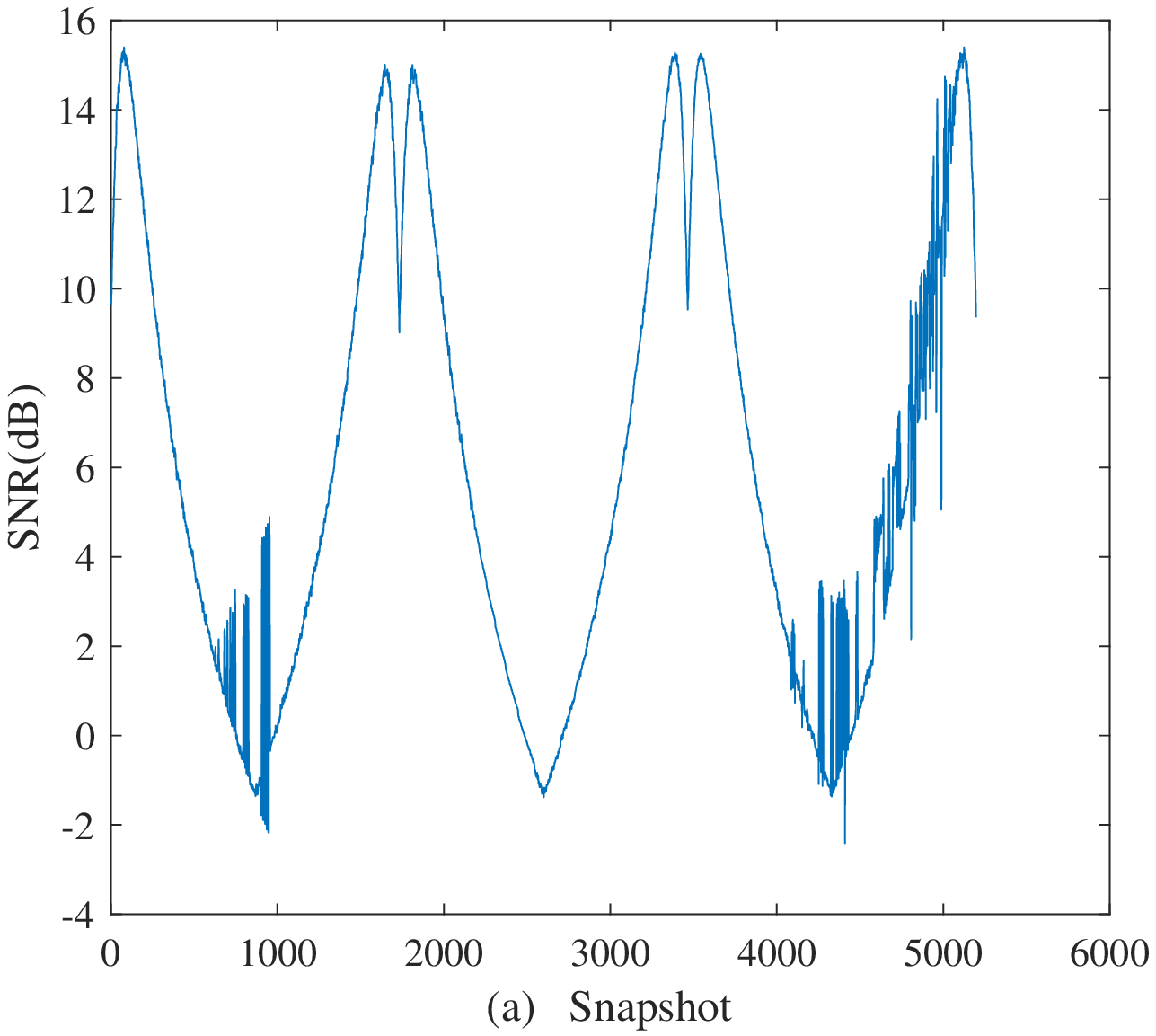} \quad
\includegraphics[width=3.5cm, height=2.8cm]{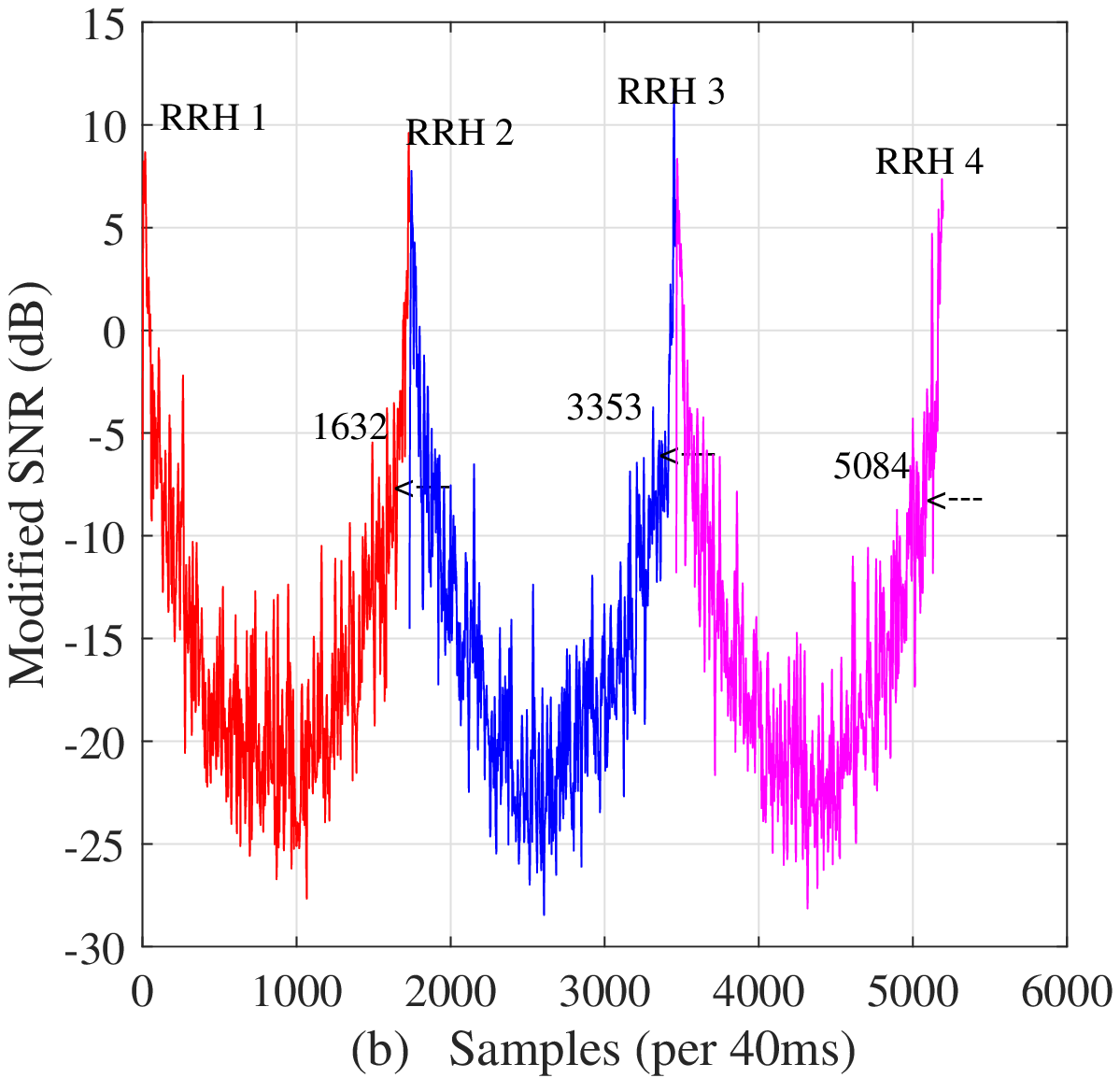} \\
\includegraphics[width=3.5cm, height=2.8cm]{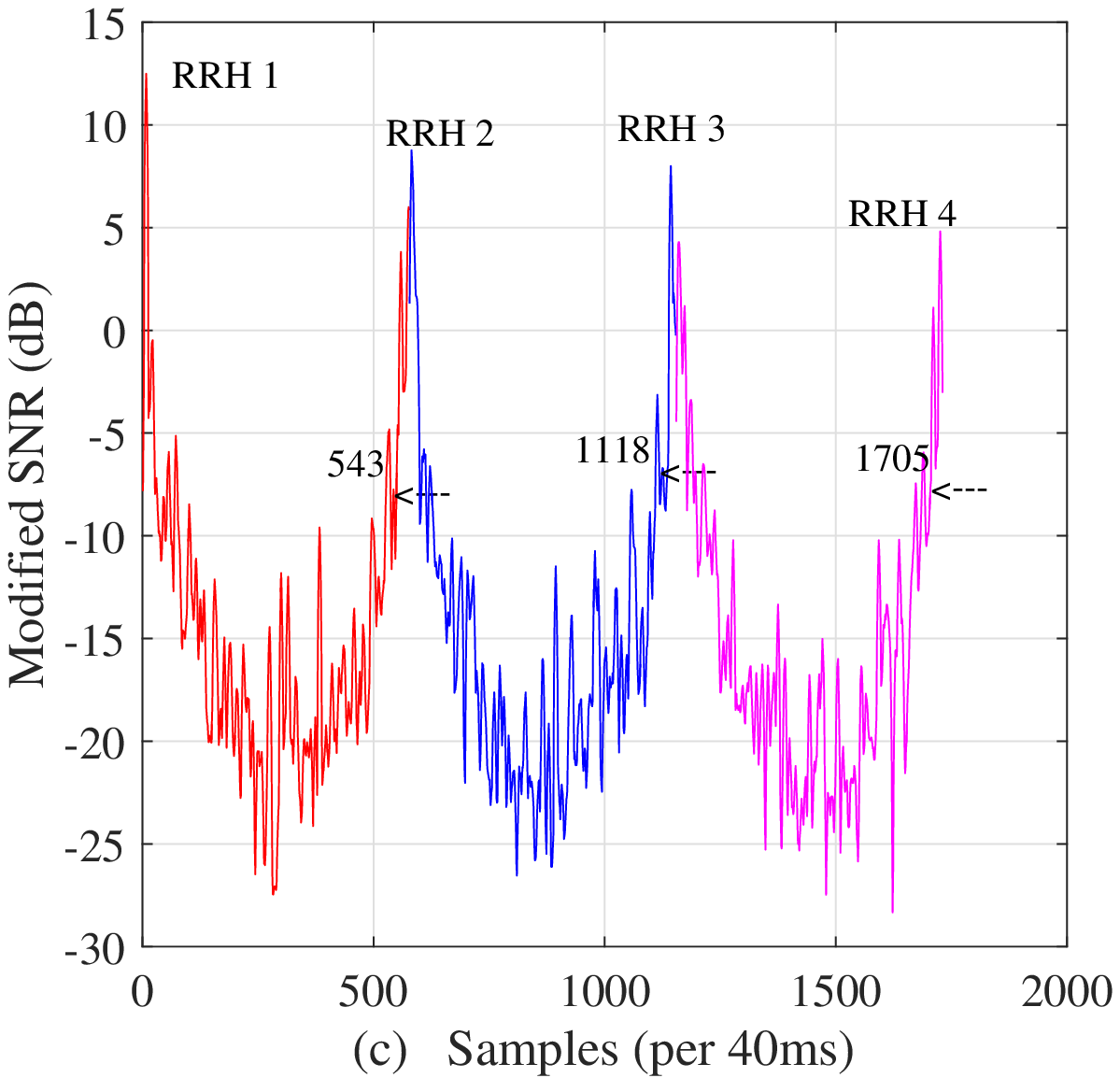} \quad
\includegraphics[width=3.5cm, height=2.8cm]{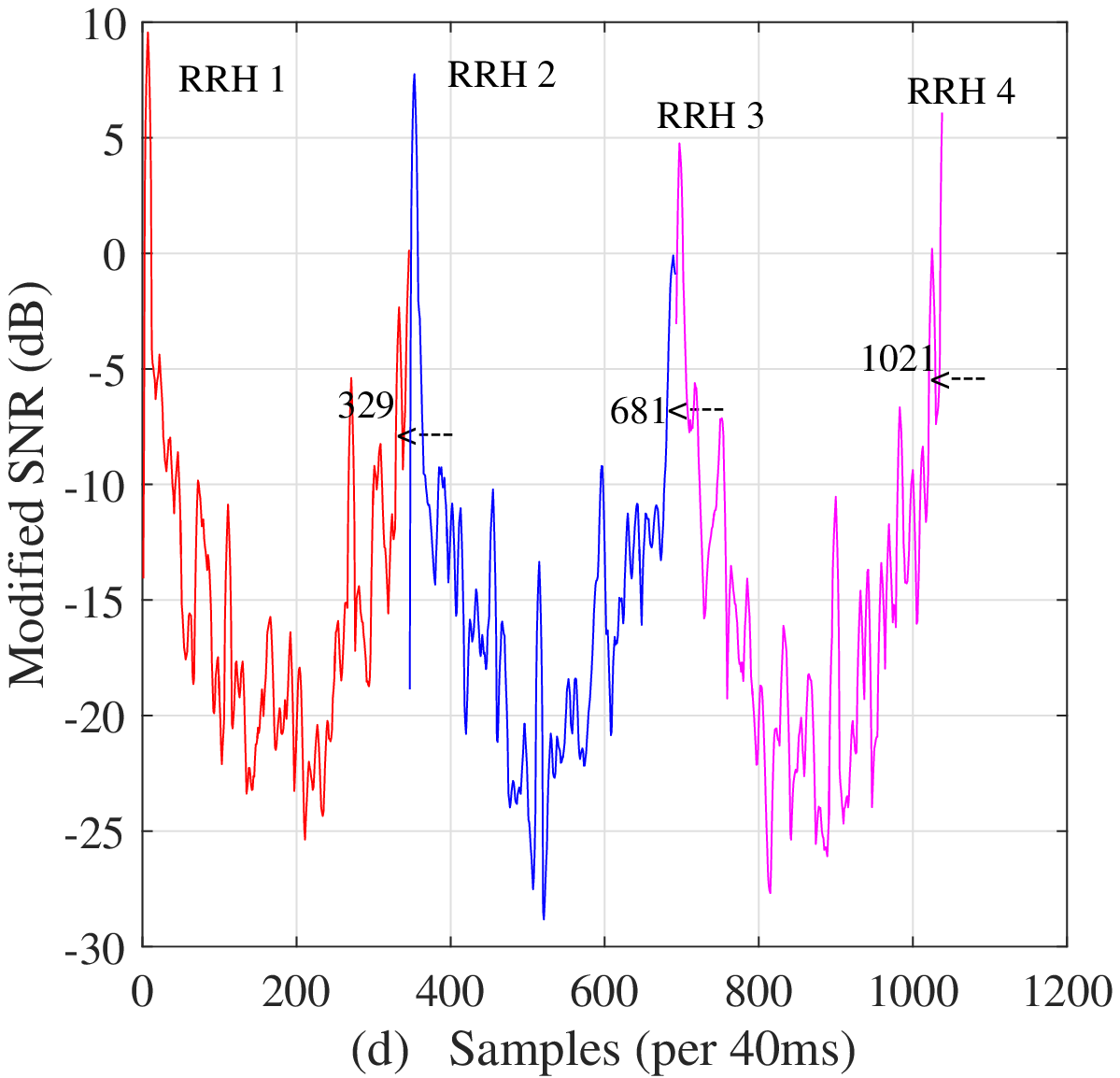} \\
\caption{SNR versus snapshot and modified SNR versus samples.}
\label{fig3} 
\end{figure}

\begin{figure}
\centering
\subfigure[Throughput versus snapshot]{
\begin{minipage}[b]{0.5\textwidth}
\includegraphics[width=1\textwidth,height = 3.5cm]{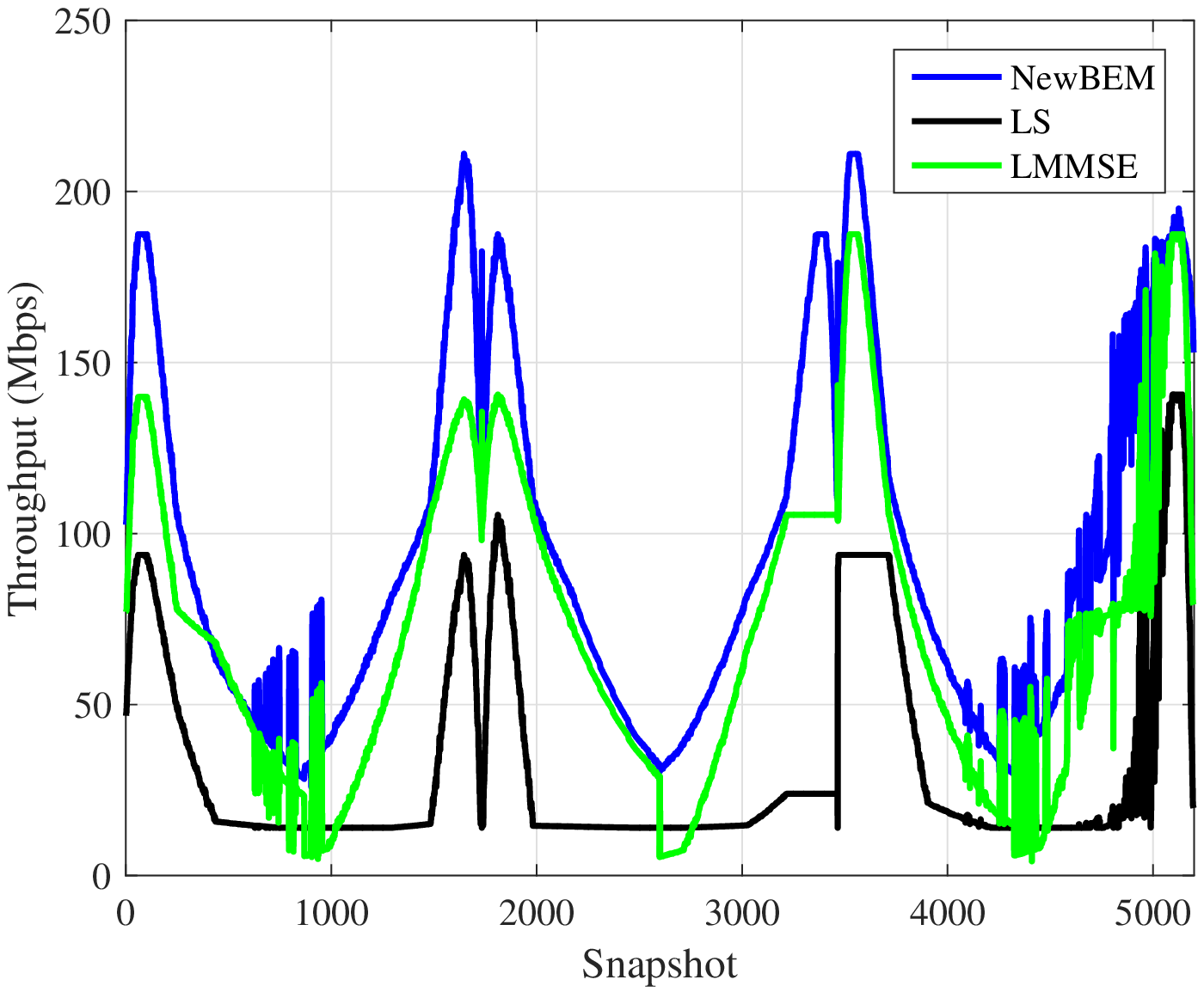}
\end{minipage}
}

\subfigure[Throughput versus samples]{
\begin{minipage}[b]{0.5\textwidth}
\includegraphics[width=1\textwidth,height = 3.5cm]{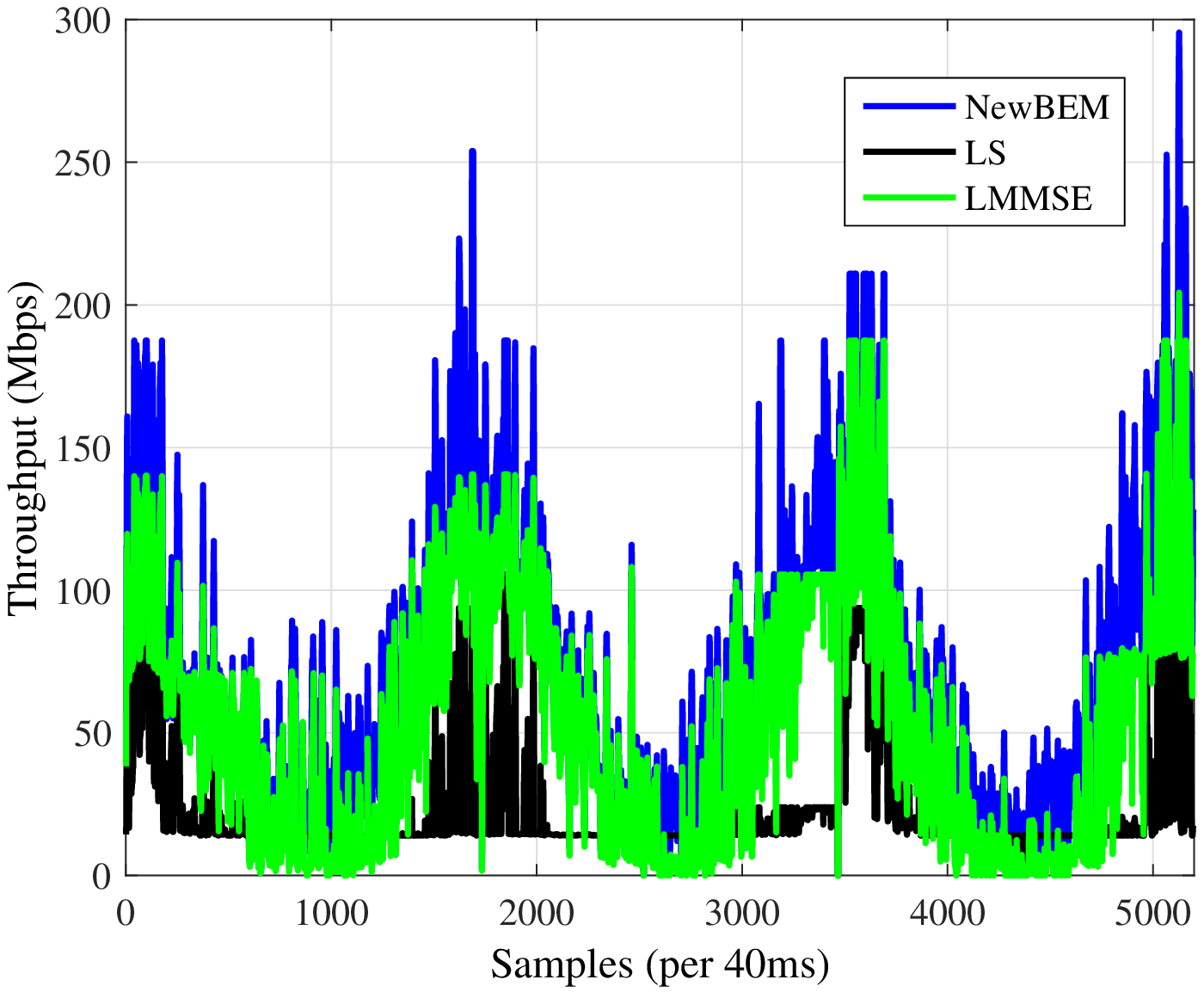}
\end{minipage}
}
 \caption{Throughput curves with 100km/h} \label{fig4}
\end{figure}

\section{NUMERICAL RESULTS AND DISCUSSION}
In this section, we provide some numerical results to discuss the handover performance. We firstly get the data of the average receiver power at a UE per meter from the simulation platform offered by the State Key Laboratory of Rail Traffic Control and Safety in Beijing Jiaotong University. These data are obtained from viaduct environment, cutting environment and urban area, respectively, and the parameters from this simulation platform are listed in Table \ref{table1}. These data are obtained by considering path loss and small-scale fading Then, we show other parameters in this section as Table \ref{table2}. From Table \ref{table2}, it is observed that we analyse the handover starting time and the total delay via different speed of HST (i.e., 100km/h, 300km/h, 500km/h).

\begin{table}[h]
\small
\centering
\caption{SimulationPlatform Parameters}
\begin{tabular}{ | l | l |}
\hline
\rowcolor[gray]{.75}
\textbf {Parameters} & \textbf{Values} \\
\hline
\rowcolor[gray]{.9}
Frequency & 3.5GHz \\
\rowcolor[gray]{.8}
Bandwidth & 100MHz	 \\
\rowcolor[gray]{.9}
Distance between RRH and railway track & 100m \\
\rowcolor[gray]{.8}
Distance between RRHs & 1732m\\
\rowcolor[gray]{.9}
Simulation length & 1732*3m\\
\rowcolor[gray]{.8}
Height of RRH & 30m\\
\rowcolor[gray]{.9}
Orientation of RRH & Turn main lobe \\
\rowcolor[gray]{.9}
 & along the track\\
 \rowcolor[gray]{.8}
RRH antenna & TX power 30dBm\\
\rowcolor[gray]{.9}
UE antenna & TX power 23dBm\\
\rowcolor[gray]{.8}
Outdoor to indoor penetration & 20dB\\
\rowcolor[gray]{.9}
Distance sample interval & 1m\\
\hline
\end{tabular}
\label{table1}
\end{table}

\begin{table}[h]
\small
\centering
\caption{Other Simulation Parameters}
\begin{tabular}{ | l || l |}
\hline
\rowcolor[gray]{.75}
\textbf {Parameters} & \textbf{Values} \\
\hline
\rowcolor[gray]{.9}
$T_s$ & 1ms \\
\rowcolor[gray]{.8}
Speed & 100km/h \quad 300km/h  \quad 500km/h\\
\rowcolor[gray]{.9}
$\sigma$ & 6dB\\
\rowcolor[gray]{.8}
$\alpha_1$ & 1/2\\
\rowcolor[gray]{.9}
$\alpha_2$ & 3/8\\
\rowcolor[gray]{.8}
$TTT$ & 40ms\\
\hline
\end{tabular}
\label{table2}
\end{table}

\subsection{SNR versus samples (i.e., per 40ms)}
Fig. \ref{fig3} shows four subfigures about SNR versus snapshot and modified SNR versus samples. As illustrated in subfigure (a), it plots SNR versus snapshot based on the average receiver power without taking $P_{ICI}$ and shadowing into consideration. Taking shadowing and $P_{ICI}$ into consideration, subfigure (b), (c) and (d) plot modified SNR versus samples with 100km/h, 300km/h and 500km/h, perspectively. Compared with subfigure (a), SNR in subfigure (b), (c) and (d) decreases at each sample because ICI and shadowing degrade the RSS.

From Section III, we know that the samples should be taken every 40ms. In subfigure (b), we take sample every snapshot due to the speed of 100km/h. In subfigure (c), we take sample every three snapshot due to the speed of 300km/h and in subfigure (d), we take sample every five snapshot due to the speed of 500km/h. 

We take an example of once handover procedure with different speed, as shown in subfigure (b), (c) and (d). The numbers marked in these three subfigures present the sample value related to the handover starting time, which can be calculated by the numbers multiply 40ms. And in these three subfigures, red curves represent the viaduct environment, while blue curves and pink curves describe cutting environment and urban area, respectively.

\begin{figure}
\centering
\subfigure[Throughput versus snapshot]{
\begin{minipage}[b]{0.5\textwidth}
\includegraphics[width=1\textwidth,height = 3.5cm]{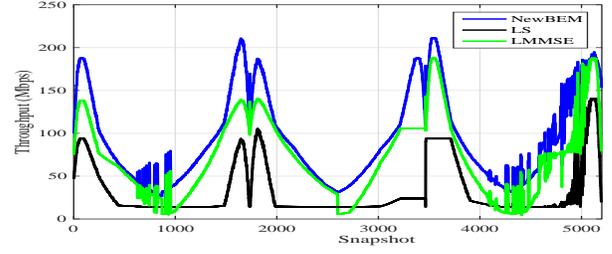}
\end{minipage}
}

\subfigure[Throughput versus samples]{
\begin{minipage}[b]{0.5\textwidth}
\includegraphics[width=1\textwidth,height = 3.5cm]{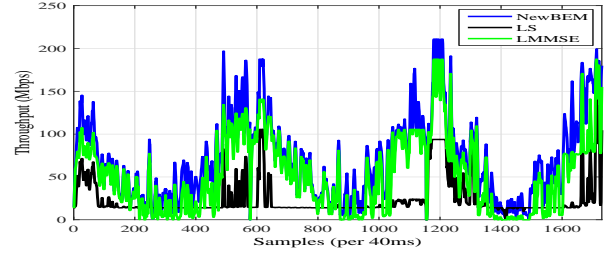}
\end{minipage}
}
 \caption{Throughput curves with 300km/h} \label{fig5}
\end{figure}

\begin{figure}
\centering
\subfigure[Throughput versus snapshot]{
\begin{minipage}[b]{0.5\textwidth}
\includegraphics[width=1\textwidth,height = 3.5cm]{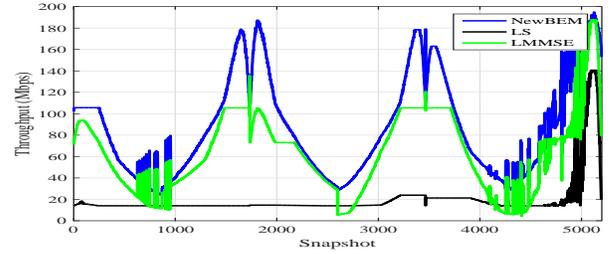}
\end{minipage}
}

\subfigure[Throughput versus samples]{
\begin{minipage}[b]{0.5\textwidth}
\includegraphics[width=1\textwidth,height = 3.5cm]{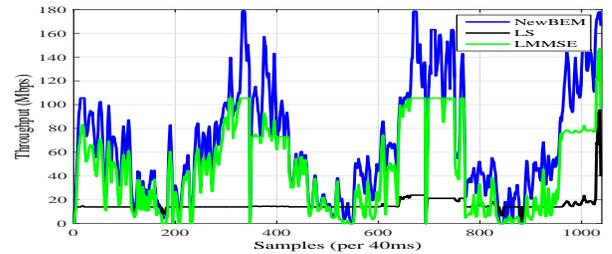}
\end{minipage}
}
 \caption{Throughput curves with 500km/h} \label{fig6}
\end{figure}

\subsection{System throughput versus samples}
Subfigure (a) in Fig. \ref{fig4}, Fig. \ref{fig5} and Fig. \ref{fig6} plot the system throughput without shadow fading and $P_{ICI}$ versus snapshot with speed 100km/h, 300km/h and 500km/h, respectively. Meanwhile, subfigure (b) in Fig. \ref{fig4}, Fig. \ref{fig5} and Fig. \ref{fig6} plot the system throughput with shadow fading and $P_{ICI}$ versus samples with speed 100km/h, 300km/h and 500km/h, respectively. The results show the throughput curves with different channel estimation techniques by taking different speed based on the simulation platform named Monkey, which is offered by Nokia. This platform supports the whole transmission process, including signal coding, channel coding, channel estimation and so on. In these figures, the channel estimation techniques are least-squares (LS) channel estimation, linear minimum mean square error (LMMSE) channel estimation and a new basis expansion model (BEM) \cite{X. Wang}. And it is observed that the new BEM shows the best performance, because it stores and analyses the past received signals to exploit the channel estimators of time dimension.

As we all known, the Doppler frequency spread becomes greater with increment of speed. In high mobility environment, the great Doppler frequency spread affects the accuracy of channel estimation, coding and etc.. Therefore, by taking the Doppler correction, subfigure (a) in Fig. \ref{fig4}, subfigure (a) in Fig. \ref{fig5} and subfigure (a) in Fig. \ref{fig6} show the throughput versus snapshot with different speed. It is observed that the influence of speed is eliminated. Besides, the new BEM channel estimation shows the best performance compared with LS and LMMSE. Similarly, subfigure (b) in Fig. \ref{fig4}, subfigure (b) in Fig. \ref{fig5} and subfigure (b) in Fig. \ref{fig6} show the throughput versus samples with different speed. Compared all subfigure (b) to its related subfigure (a), throughput decreases a little. That is because the existing of $P_{ICI}$ makes $P_r$ (i.e., RSS) smaller than before.

\begin{figure}[!t]
\centering
\includegraphics[width=0.48\textwidth,height = 5cm]{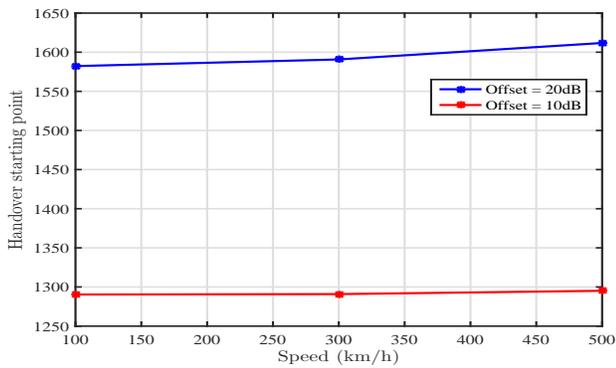}
\caption{Handover starting point versus speed.}
\label{fig7} 
\end{figure}

\subsection{Handover starting time and delay}
We simulate the handover procedure for over 500 times to discuss the handover starting point and delay. We calculate the weighted starting point and map the simulation results on the snapshot. The starting point means the distance between the handover position and the postion of the related serving RRH. As shown in Fig. \ref{fig7}, the greater offset causes the handover happens later obviously. That because the time to trigger event A3 is postponed. Moreover, the handover happens a little bit later with higher speed, which is analysed in Section III. 

By simulation, the delay is about 120ms (i.e., 3 samples) for different speed. So we conclude that the total delay is not affected by the speed.

\section{CONCLUSION}
This paper investigated the handover performance on HST via different speed under three scenarios, which are viaduct environment, cutting environment and urban area. We analysed handover procedure in detail and we also analysed the effect of high speed on handover starting time. Simulation results showed some parameters in the proposed system (i.e., SNR, throughput, handover starting time and delay).  

\small
\bibliographystyle{IEEEtran}

\end{document}